\documentclass[12pt]{article}

\usepackage[dvips]{graphicx}
\usepackage{amsmath}
\usepackage{amssymb}

\def\dd{\mathrm{d}}

\newcommand{\captive}[1]{\rule{5mm}{0mm}%
\begin{minipage}{140mm}\caption[small]{#1}\end{minipage}}
\newcommand{\captin}[1]{\rule{5mm}{0mm}%
\begin{minipage}{150mm}\caption[small]{#1}\end{minipage}}

\begin{document}
\begin{flushright}
DTP-00/37\\
\tt{hep-ph/0005208}
\end{flushright}
\vspace{6mm}
\begin{center}
{\huge{There is no $\kappa(900)$ }}\\[10mm]
{\Large{S. N. Cherry and M. R. Pennington}}\\[5mm]
{\large{\it{Centre for Particle Theory,}}}\\[1mm]
{\large{\it{University of Durham,}}}\\[1mm]
{\large{\it{Durham DH1 3LE, U.K.}}}\\[1mm]
\end{center}
\vspace{1cm}

\begin{abstract}
In the $I=0$ sector there are more scalar mesons than can fit in one
$q{\overline q}$ nonet.
Consequently, many have claimed that there is in fact more than one
multiplet, perhaps both  $q{\overline q}$ and $qq{\overline {qq}}$.
Such proposals require the existence of at least two strange
isodoublets (and their antiparticles). 
The current PDG Tables list just one state, the $K^*_0(1430)$, while
fits to data with Breit-Wigner forms and variable backgrounds can
accommodate a $\kappa(900)$ too. 
Whether a state exists in the spectrum of hadrons is not a matter of
ability to fit data along the real energy axis, but is completely
specified by the number of poles in the complex energy plane. 
Here we perform as model-independent an analytic continuation of the LASS
$\pi K$ scattering results between 825~MeV and 2~GeV as presently possible
 to determine the
number and position of resonance poles. 
We find that there {\bf is} a $K^*_0(1430)$, but {\bf no}
$\kappa(900)$.
The LASS data cannot rule out the possibility of a very low mass
$\kappa$ well below 825~MeV. 
\end{abstract}
\parskip 2mm
\baselineskip=6.65mm
\noindent{\bf PACS} numbers: 11.55.-m, 14.40.Ev

\section{Introduction}
For many years the scalar mesons have caused controversy. 
There is little consensus on the composition of many states, whilst
for some their properties and even existence is a matter for debate. 
In the latest edition of the PDG Tables~\cite{Groom}, there are four
scalar-isoscalars below 1.55~GeV, with a fifth at 1.7~GeV.
This is obviously too many for a standard $q{\overline q}$ nonet.
However, many QCD-motivated models predict the existence of non-$q
{\overline q}$ mesons, such as $qq{\overline {qq}}$ 
states~\cite{Jaffe1}, $K{\overline K}$ molecules~\cite{Weinstein} and
glueballs~\cite{glueballs}, and it is precisely in the isoscalar
sector that these unconventional mesons are most likely to be found. 
This excess of isoscalars, together with the existence of two
isovectors, the $a_0(980)$ and $a_0(1450)$, has been suggested as
evidence for two scalar nonets~\cite{Black2,Jaffe2}: the conventional
$q{\overline q}$ nonet lying around 1.4~GeV with an unconventional one
centred around 1~GeV. 
However, the PDG lists only one pair of scalar-isospinors in this mass
region, the $K^*_0(1430)$. 
This has led some authors to postulate a light strange meson, known as
the $\kappa$~\cite{Jaffe1,Scadron,Delbourgo}. 
Evidence for this resonance has been claimed within certain 
models~\cite{Black1,Ishida,VanBeveren,Oller1}, whilst other studies
dispute this~\cite{Tornqvist,Anisovich}. 
Fits to $\pi K$ production data, using Breit-Wigner forms, some with
arbitrarily varied backgrounds, others with theoretically motivated
backgrounds, allow a $\kappa$ to be accommodated. 
However, the existence of a state is not a matter merely of the
quality of fit to data along the real energy axis. 
Indeed, a state is wholly specified by there being a pole of the
$S-$matrix in the complex energy plane on the nearby unphysical
sheet.
Such specification, of course, does require experimental information in
the relevant region of the real energy axis.

Where this light strange meson has a mass around 900~MeV, it comes within
the energy domain testable with the highest statistics results on $\pi
K$ scattering, which come from the LASS experiment~\cite{Aston}. 
In this paper we present the results of a close to model independent
determination of the number and position of  poles in the $I =
\frac{1}{2}$, $J = 0 \ \pi K$ scattering amplitude based on these LASS
data above 825~MeV. 
The technique relies wholly on the analytic properties of $S-$matrix
elements and  requires no artificial separation into resonance and
background contributions. 
We also consider continuations using results from an earlier SLAC
experiment~\cite{Estabrooks}. 
In Sect.~2, we present the method and then in Sect.~3  describe the
experimental inputs we use. 
In Sect.~4, we apply the analytic continuation methodology first to
model data to assess its capabilities and limitations. 
In Sect.~5, we go on to study the analytic continuation using the real
experimental results. 
We find there is no $\kappa(900)$.
Since the method is one of continuing experimental data and the LASS
experiment only provides information above 825~MeV we cannot rule out
the possibility of an even lower mass $\kappa$. 
In Sect.~6, we conclude.

\section{Method}
The method we use is due to Nogov\'a {\it et al.}~\cite{Nogova}.
It combines simple statistics with the analytic properties of the
scattering amplitude to locate the positions of poles. 
As usual, we expect resonances to correspond to poles on unphysical
sheets~\cite{Taylor}. 

\subsection{Mapping}
\begin{figure}[!htb]
\vspace{2mm}
\begin{center}
\includegraphics[angle=270,width=0.8\textwidth]{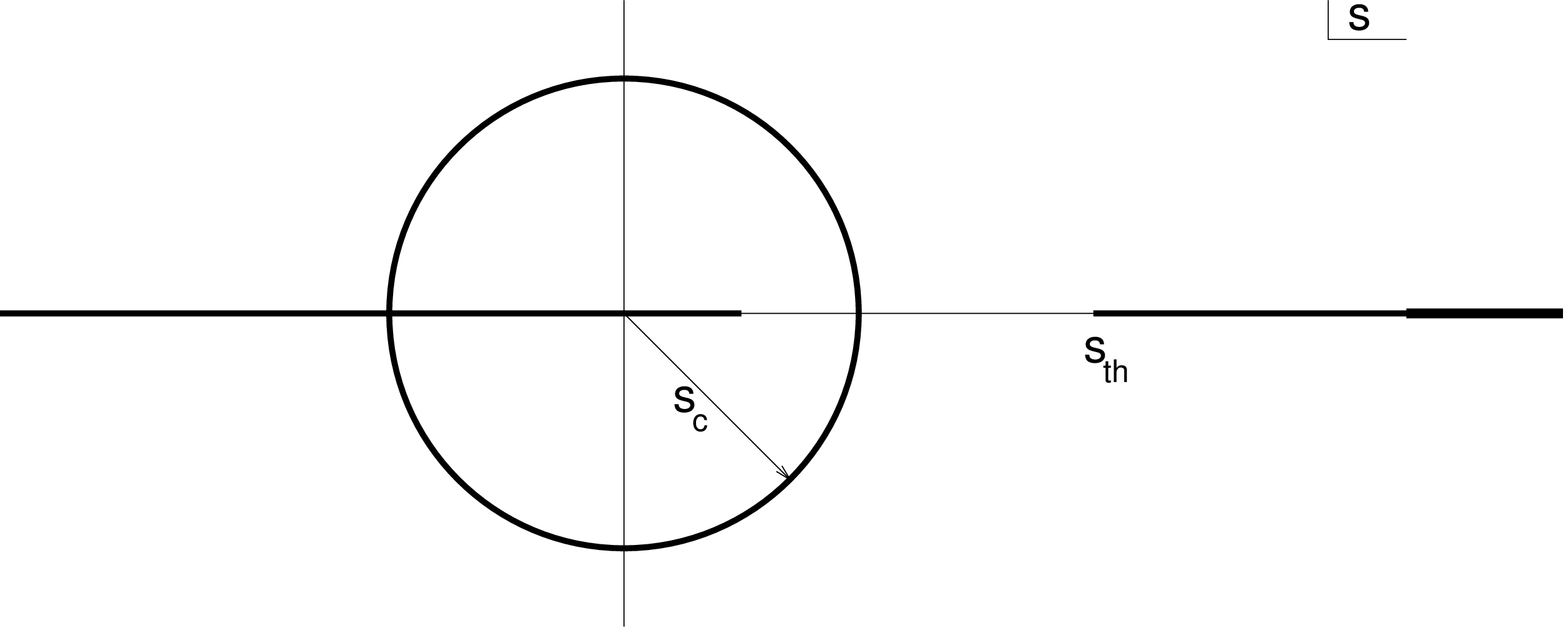}
\vspace{-6mm}
\captive{The cut structure of the $\pi K$ partial wave amplitude,
where \mbox{$s_{th}=(m_K+m_{\pi})^2$} and the radius of the circular
cut is \mbox{$s_c=m_K^2-m_{\pi}^2$}.\label{scut}} 
\end{center}
\vspace{-3mm}
\end{figure}

In order to locate poles in the complex plane from scattering data on
partial wave amplitudes along the real axis, we must perform an
analytic continuation. 
To maximise the domain in which this continuation is valid, we begin
by conformally mapping the partial wave amplitude into the unit disc.
As shown in Fig.~\ref{scut}, for $\pi K$ scattering the $s-$plane is
split into two regions and in this case we map the region
\emph{outside} the circular cut.
The mapping is designed so that the cuts of the partial wave amplitude
in the $s-$plane are mapped onto the circumference of the circle in
the $z-$plane.
This allows us to continue the amplitude throughout the complex $z-$plane.

\begin{figure}[!htb]
\vspace{5mm}
\begin{center}
\includegraphics[width=12cm]{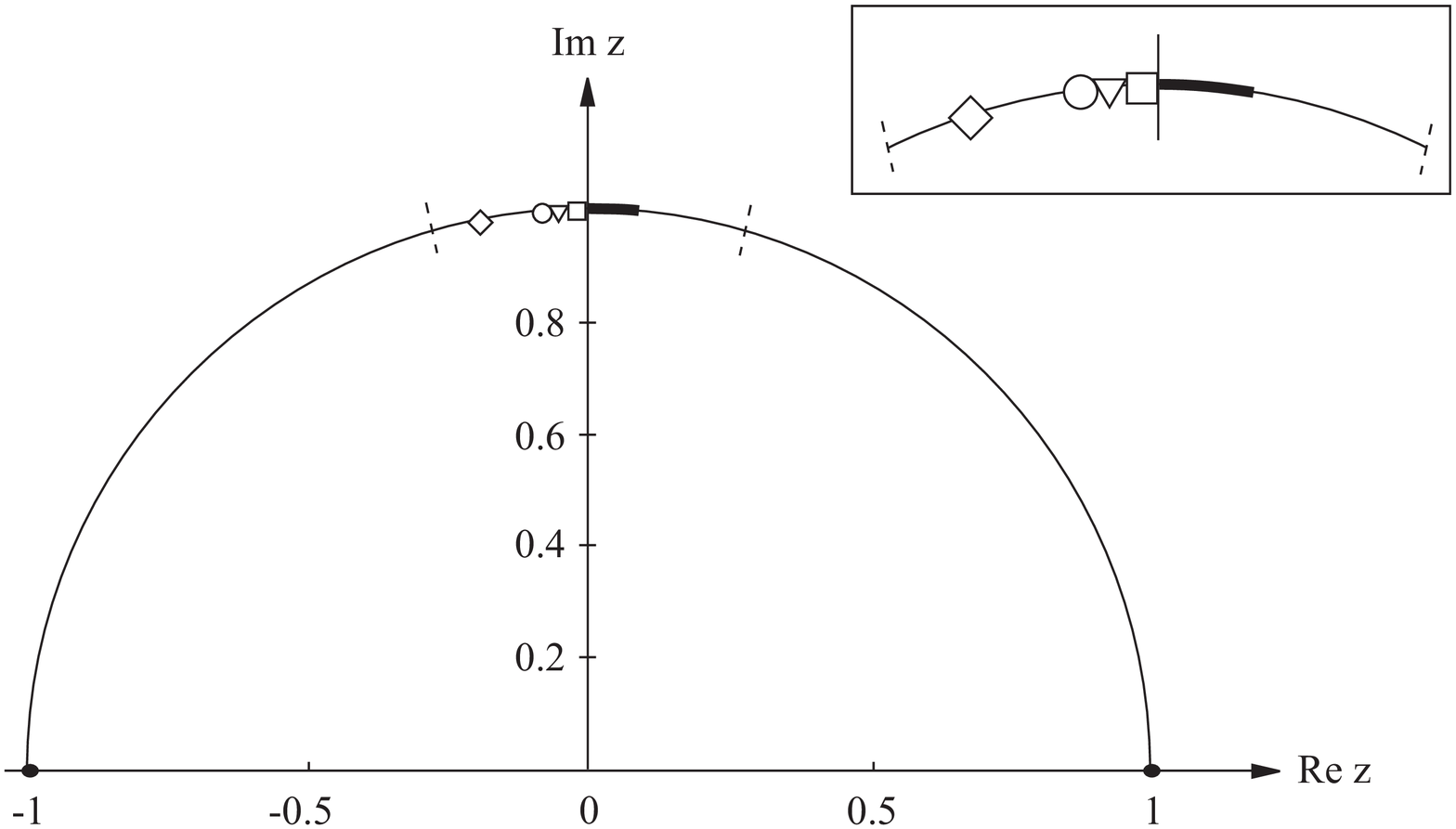}
\captive{The z-plane showing how points in s-plane map.
The mapping parameter $\beta$ is chosen so that
\mbox{$s=(1.4+0.15i)^2$~GeV$^2$} is mapped to the imaginary axis. 
The thicker line shows the arc covered by the LASS data.
The inset shows an enlargement of the key region close to $z\,=\,i$,
from where we analytically continue. 
The symbols mark particular values of $s$ as follows: $
\blacklozenge~s~=~s_{th},\ \square~s~=~\infty,\  \triangledown \, s =
-s_c,\   \circ \, s = i s_c,\  \lozenge \, s = \frac{1}{\sqrt{2}}(s_c
+ i s_c),\  \bullet \, s = s_c$. 
\label{zcut}}
\end{center}
\vspace{-3mm}
\end{figure}

The mapping is accomplished in two steps.
Firstly, 
\begin{equation}
y(s) = \left( \frac{s-s_c}{s+s_c} \right) ^2\qquad ,
\label{map1}
\end{equation}

\noindent which maps the real cuts in the $s-$plane onto the positive
real axis of the $y-$plane and the circular cut onto the negative real
axis. 
Then to map the twice cut $y-$plane into the unit disc, we define
\begin{equation}
z(s) = \frac{i \beta \sqrt{y(s)} - \sqrt{y(s)-y(s_{th})}}{i \beta
\sqrt{y(s)} + \sqrt{y(s)-y(s_{th})}}\quad , 
\label{map2}
\end{equation}

\noindent  where $\beta$ is a real parameter which is chosen so that
the region of interest in the $s-$plane is mapped close to the
imaginary axis in the $z-$plane, thereby minimising the distance the
continuation must cover. 
The points $s=s_{th}$ and $s=s_c$ are fixed for any value of $\beta$,
being mapped to $z=1$ and $z=-1$ respectively. 
Notice that the cuts in the upper half of the $s-$plane (such as the
region where physical data lie) are mapped to the upper semi-circle in
Fig.~\ref{zcut}. 

It is obvious from this mapping that the physical region in the
$s-$plane only covers a fraction of the circle. 
Even if we had data over an infinite range of energies, we could never
complete the circle, see Fig.~\ref{zcut}. 
Nevertheless, we only need analytically continue very close to the
region of data (i.e. the solid arc in Fig.~\ref{zcut}). 
We shall return to this point later.

From Fig.~\ref{zcut} we can see that the mapping is highly non-linear
in the following sense. 
As we increase the energy from threshold to infinity, the proportion
of the circle covered by each increment falls very sharply. 
It is clear that the region between threshold (at 633~MeV) and the
start of the data (at 825~MeV) covers a much longer arc than the
region in which we have data (825~MeV to 2.51~GeV) and the region
between the end of the data and infinite energy is much smaller
still. 
This compression of the high energy region would suggest that the
method is less sensitive to higher mass resonances. 
This apparent weakness actually has some benefit.
As we explain in the next section, we test the scattering amplitude
against the hypothesis that it contains a given number of poles. 
The computational burden increases rapidly as we increase the number
of poles, so if we had to account for all the possible radial
excitations in a given channel the calculation would soon become
computationally prohibitive. 
Consequently, the procedure is only practicable for the lowest few states.

\baselineskip=6.37mm

\subsection{Analytic Continuation}
For now we assume that we have a scattering amplitude, with errors,
$\Delta_i$,\footnote{This method requires the errors on the real and
imaginary parts of the amplitude to be equal. 
This simplification is arranged by taking the value of the error to be
the larger of the error on the real part and the error on the
imaginary part at any given energy.}, defined at discrete points all
the way around our circle $\vert z \vert = 1$.
We define $\vert \delta z_i \vert$ as the average distance between the
$i$th data point and its two nearest neighbours in the complex
$z-$plane shown in Fig.~\ref{zcut}.
In regions where these discrete points are densely packed, our
scattering amplitude is most tightly controlled and so we weight the
error on each point by the density of the data points in that region,
$\epsilon_i(z) = \Delta_i \sqrt{\vert \delta z_i \vert/2 \pi}$. 
We make a smooth interpolation to the amplitude, $Y(z)$, and the
weighted errors, $\epsilon(z)$, so that we have continuous functions
defined on the entire circle.  
This weighting procedure ensures that the small region of the unit
circle in Fig.~\ref{zcut}, where we have experimental data, controls
the analytic continuation into the nearby region where resonance poles
are expected to sit. 
If $F(z)$ is a square integrable function on the circle, $C$, then we
can test how well this function describes the data through a $\chi^2$,
defined by 
\begin{equation}
\chi^2\, =\, \frac{1}{2\pi} \oint_C \bigg\vert \frac{F(z) -
Y(z)}{\epsilon(z)}\bigg\vert^2 \, \vert \dd z \vert \label{chi1}
\quad. 
\end{equation}

We now introduce a non-zero weight function $g(z)$, which is defined
to be real analytic and constrained by $\vert g(z) \vert = \epsilon
(z)$ around the circle. 
We expand the data and the trial function as Laurent series about the
origin, so that 
\begin{equation}
y(z) = \frac{Y(z)}{g(z)} = \sum_{k = -\infty}^{\infty} y_k\, z^k \qquad ,\qquad
f(z) = \frac{F(z)}{g(z)} = \sum_{k = -\infty}^{\infty} a_k\, z^k\quad . \label{ser}
\end{equation}

\noindent Since partial wave amplitudes are real analytic, the
coefficients $a_k$ and $y_k$ are real. 
Although we are expanding about the origin, the expansion is carried
out round the circle, since this is where we have data. 
The singular Laurent coefficients will pick up any poles within the
disc. 
Substituting Eq.~(\ref{ser}) into Eq.~(\ref{chi1}) gives
\begin{equation}
\chi^2 = \sum_{k = 1}^{\infty} \left(a_{-k} - y_{-k}\right)^2 +
\sum_{k = 0}^{\infty} \left(a_k - y_k\right)^2\quad . \label{chi2} 
\end{equation}

\noindent The pole structure of our partial wave amplitude, $Y(z)$,
can then be determined by finding the test function $F(z)$ which
minimises the first summation in Eq.~(\ref{chi2}). 

If we want to test against the assumption that there are no poles in
the data we must use a test function that is analytic, i.e. $a_{-k}
\equiv 0 \ \textrm{for } k > 0$. 
Then, if the  amplitude has no poles, the quantity $\chi^{\,2}_{0} =
\sum_{k=1}^N y_{-k}^{\,2}$ will be zero \footnote{Actually, due to
inevitable experimental noise, $\chi^{\,2}_{0}$ will not be exactly
zero, but should fit a $\chi^2$ distribution with $N$ degrees of
freedom.}.
If we think that the scattering data has one pair of complex conjugate
poles in the $z-$plane (as is the case where there is one resonance
present) then we can write our test function as  
\begin{eqnarray}
f(z) &  = & \frac{\alpha}{z-z_0}\, +\, \frac{\alpha ^*}{z-z_0^*}\, +\, h(z) \nonumber \\
&  = & \sum_{k=1}^{\infty} \frac {2\, \Re[\alpha z_0^{k-1}]}{z^k}\, +\, h(z) \quad ,\label{Q1}
\end{eqnarray}

\noindent where $h(z)$ is some analytic function.
By comparing Eq.~(\ref{Q1}) with Eq.~(\ref{ser}) we see that, in this
case, the analytic continuation is carried out by setting $a_{-k} =
2\, \Re[\alpha z_0^{k-1}]$ and then it is the quantity $\chi^2_1 =
\sum_{k=1}^N (y_{-k} -  2\, \Re[\alpha z_0^{k-1}])^2$ that has a
$\chi^2$ distribution with $N$ degrees of freedom as long as $\vert
z_0 \vert \ll 1$. 

However, in practice $\vert z_0 \vert$ is close to one and the above
procedure becomes unreliable. 
Consequently, we adopt the alternative approach of cancelling any pole
in the data explicitly by the introduction of the so-called Blaschke
pole-killing factor.
This avoids the need to make the expansion explicit in Eq.~(\ref{Q1})
and so significantly improves the convergence. 
For a pair of complex conjugate poles the Blaschke factor looks like
\begin{equation}
B_{z_0} = \frac{(z-z_0)(z-z_0^*)}{(1-z z_0)(1-z z_0^*)}\quad .
\end{equation}

\noindent We define the function 
\begin{equation}
\tilde{y} (z) = \frac{Y(z) B_{z_0} (z)}{g(z)} =
\sum_{k=-\infty}^{\infty} \tilde{y}_{k}\, z^k 
\end{equation}

\noindent and then the quantity $\sum_{k=1}^N (\tilde{y}_{-k})^2$ will
indeed obey a $\chi^2$ distribution with $N$ degrees of freedom, if
the actual amplitude contains one pair of complex conjugate poles. 

\section{Experimental Input}
Our experimental information on the $S-$wave $\pi K$ partial wave
amplitude comes in the form of the magnitudes, $a(s)$, and phases,
$\phi(s)$, measured, for instance, by the LASS experiment.
The partial wave amplitude on sheet I is normalised thus
\begin{equation}
f^I(s) = \frac{a(s)\, e^{i\phi(s)}}{\rho (s)}\qquad ,
\end{equation}

\noindent where $\rho(s)= 2q/\sqrt{s}$ and  $q$, the c.m. 3-momentum
\begin{equation}
q =\,\frac{1}{2}\, \sqrt{\frac{[s-(m_K + m_{\pi})^2][s-(m_K -
m_{\pi})^2]}{s}}\quad . \label{mom} 
\end{equation}

$\pi K$ scattering has two possible isospin channels, $I = 1/2$
and $I = 3/2$. 
The resonances we consider are isodoublets, as there are no known
resonances in the $I = 3/2$ channel. 
This means that there is no need to separate the isospin components to
determine the pole positions, whereas this is essential in a
Breit-Wigner-type fit. 
We therefore carry out our analysis for the total $S-$wave data.
However, as a check we also perform this for the  $I = 1/2$
results alone. 
Of course, the  $I = 1/2$ data must be extracted from the total
$S-$wave data by subtracting an assumed form for the  $I = 3/2$
amplitude, thereby introducing  an additional layer of uncertainty. 
Moreover, this separation is only possible over a limited energy range.
Nevertheless, the $I = 1/2$ component does provide a valuable
check. 

As is well-known, resonance poles do not appear on the physical sheet
in the energy plane. 
Consequently, we must move to the relevant unphysical sheet.
For purely elastic scattering this would be sheet II. In practice for
$\pi K$ scattering, the $\eta K$ channel opens very weakly (in
agreement with $SU(3)_F$ expectations)  and so any inelasticity can be
safely neglected, until one reaches the $\eta^{\prime} K$
threshold. This channel opens in the region of a possible
$K_0^*(1430)$  and complicates the sheet structure. 
Any resonance would then be on sheet III\footnote{The Riemann sheets
are labelled by $(r_1,r_2)$, where the $r_j$ are the signs of the
imaginary parts of the complex phase spaces $\rho_j$ in particular
channels. 
$j=1$ for $\pi K$, $j=2$ for $\eta^{\prime} K$.
\newline Sheet I is $(+,+)$, sheet II is $(-,+)$ and sheet III is
$(-,-)$}. 
We must take this threshold into account when we change sheets. 
The simplest way to pick the correct sheet is to change the sign of
the phase, i.e. 
\begin{equation}
f^{II}(s) = \frac{a(s)\, e^{-i \phi(s)}}{\rho(s)}\qquad .
\end{equation}

\noindent This moves us onto sheet II below the $\eta^{\prime} K$
threshold and sheet III above it. 
This method of changing sheets is also valid for the total $S-$wave
amplitude. 
We then map the data, including errors, as described above.

As noted previously it is not possible to cover the circle completely
with physical region data, since we do not have experimental results
either close to threshold or on the circular and left hand cuts (see
Fig.~\ref{scut}).
The exact proportion of the circle covered by the direct channel $\pi
K$ results depends on the choice of the mapping parameter $\beta$.
How we treat the near threshold arc of Fig.~\ref{zcut} is a little
different for our model tests of Sect.~4 and when we consider the
actual experimental data in Sect.~5.
Consequently, we detail their treatment in the appropriate section.
As we do not have information about what happens on the unphysical
cuts, we simply make a guess. 
So that the guess does not unduly affect the results, we de-weight
these guessed points by giving them very large errors and by ensuring
that they are widely spaced (see Fig.~\ref{data}).
With the top semi-circle now spanned by data points, we make an
interpolation to give us a continuous function. 
We complete the circle by reflecting the interpolated data in the
upper semi-circle onto the lower half so as to obey the Schwarz
Reflection Principle. 

A suitable form for the weight function $g(z)$, which fulfills all the
conditions described above is 
\begin{equation}
g(z)\,=\, \exp \, \left(\sum_{n=0}^{N} c_n z^n\right)
\end{equation}

\noindent where the $c_n$ are found from a Fourier cosine fit to $\ln
\epsilon(z)$. 
We take $N=100$.
The singular coefficients of the Laurent expansion of the data about
the origin are just 
\begin{equation}
y_{-k}\, =\, \frac{1}{2 \pi} \oint_c \frac{Y(z)\, z^k}{g(z)} \  \vert
\dd z \vert\qquad . 
\end{equation}
\vspace{1.5mm}

\section{Results from Model Data}
As a first application, the method described in Sect.~3 is tested on a
model amplitude describing the situation where a light, broad
resonance, $\kappa_1$, and a heavier, narrower resonance, $\kappa_2$,
are present\footnote{This generalises a model analysis discussed by
Nogov\'a {\it et al.} to two resonances.}. 
The resonances are constructed using Jost functions, with the lighter,
broader resonance being treated as the background at the second
resonance, {\it i.e.}
\begin{equation}
f^I (s) = f^I_1(s) + e^{2 i \delta_1(s)} f^I_2(s)
\end{equation}
on the first sheet, where
\begin{eqnarray}
f^I_j (s) &=& \frac{-2 k d_j}{\rho\, (k + c_j + i d_j)(k - c_j + i
d_j)}\quad ,\\ 
\delta_1(s) &=& \arctan \left(\frac{2 k d_1}{c_1^2 + d_1^2 - k^2}
\right)\quad ,\\ 
k &=& \frac{1}{2}\sqrt{s-(m_K + m_{\pi})^2}\quad .
\end{eqnarray}

\begin{figure}[!htb]
\begin{center}
\includegraphics[width=0.49\textwidth,height=0.35\textheight]{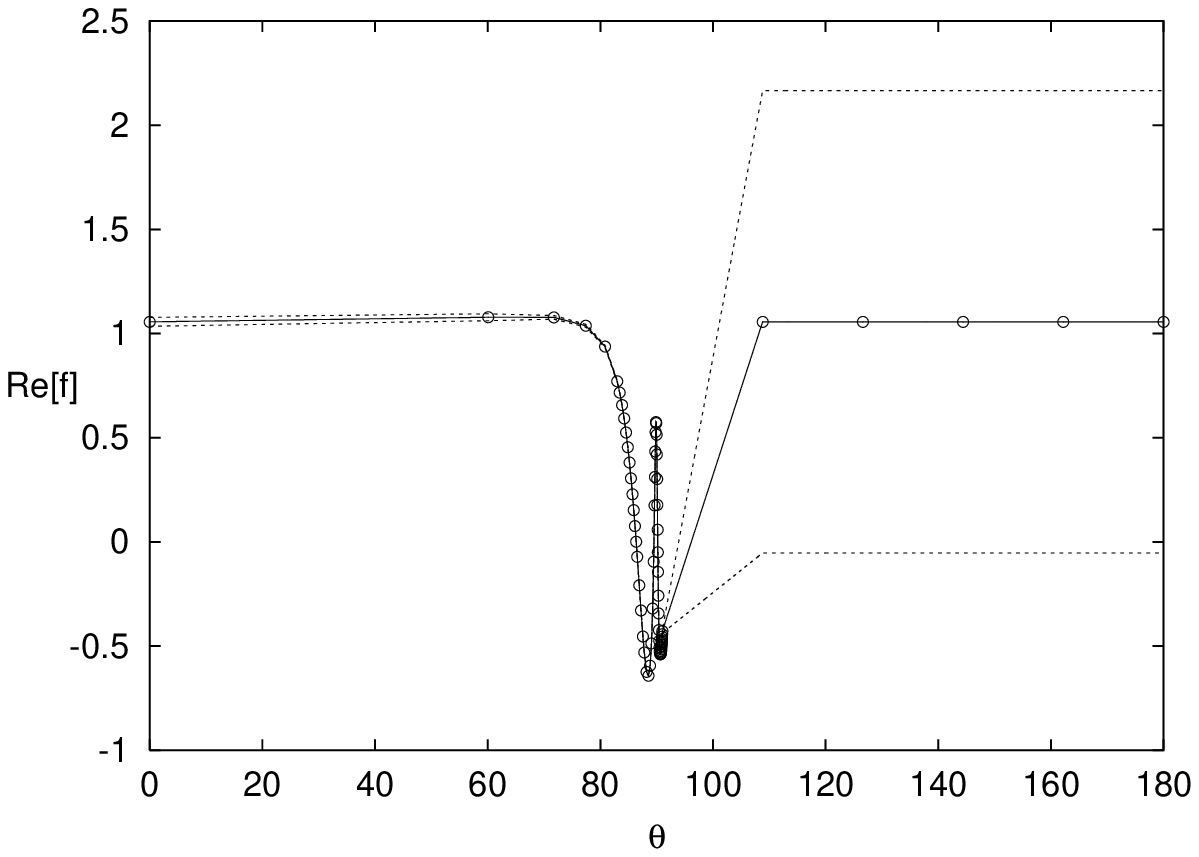}
\includegraphics[width=0.49\textwidth,height=0.35\textheight]{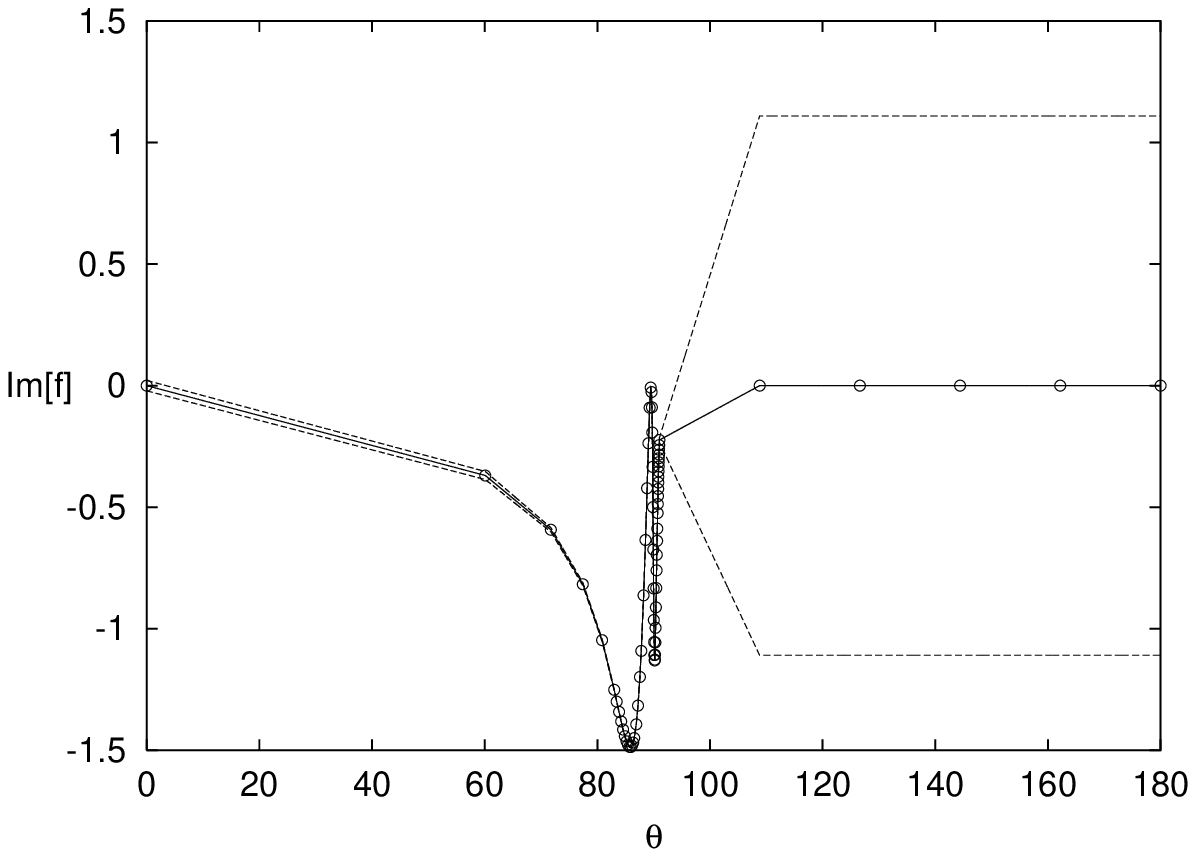}
\captive{Real and imaginary parts of $f^{II}$ as a function of
$\theta$, the angular position around the circle in degrees, for the
model data-set. 
The circles represent data points, the solid line shows the
interpolation to the data and the dashed line shows the interpolated
weighted errors. 
\label{data}}
\end{center}
\vspace{-5mm}
\end{figure}

\begin{sloppypar}
\noindent The four parameters $\left\{c_j,d_j \right\}$ are chosen, so
that the amplitude has poles at \mbox{$s_1 = (0.9 \pm 0.25 i)^2 \
({\rm GeV})^2$} and \mbox{$s_2 = (1.4 \pm 0.15 i)^2 \
({\rm GeV})^2$}. 
Data are created for energies equivalent to the full LASS
data-set~\cite{Aston}, with the error on both the magnitude and phase
fixed at 5\%. 
Below the energy of the LASS data, we create a few equally spaced
points between threshold and and 825~MeV. 
This will be discussed in more detail later. 
The amplitude in the unphysical region is set to a real constant
(incidentally equal to the amplitude at threshold) with very large
errors ($\pm 5$). These model data are illustrated in
Fig.~\ref{data}. 
\end{sloppypar}

These model data allowed us to verify that the Blaschke factor method
is indeed more stable to variations of non-physical parameters (such
as the number of terms in our definitions of the $\chi^2$ and the
weight function) and hence more accurate. 
The pole positions and $\chi^2$'s found by fitting over the first 40
singular coefficients using this procedure are shown in Table~\ref{TJR}.  
Also tabulated for comparison are the results obtained when the errors
on the points in the unphysical region are halved. 

\begin{table}[!htb]
\vspace{4mm}
\begin{center}
\begin{tabular}{|c|c|c|c|r|}
\hline
Option & No. of & $z_{pole}$ & $\sqrt{s_{pole}}$ & $\chi^2$~~ \\
& resonances & $(r,\theta)$ & (MeV) & \\
\hline \hline
& 0 & --- & --- & 4081~~ \\
\cline{2-5}
1 & 1 & (0.988,89.31$^o$) & $1201 \pm 131 i$ & 281~~ \\
\cline{2-5}
& 2 & (0.993,89.98$^o$) & $1396 \pm 142 i$ & 0.5\\
& & (0.943,88.64$^o$) & $903 \pm 234 i$ & \\
\hline
& 0 & --- & --- & 3759~~ \\
\cline{2-5}
2 & 1 & (0.989,89.13$^o$) & $1173 \pm 111 i$ & 417~~ \\
\cline{2-5}
& 2 & (0.993,89.97$^o$) & $1392 \pm 140 i$ & 1.0\\
& & (0.942,88.61$^o$) & $900 \pm 233 i$ &\\
\hline
\end{tabular}
\captin{Pole positions and $\chi^2$'s for model data. 
Option 1 has the unphysical errors set to 5.
Option 2 has the unphysical errors set to 2.5.
\label{TJR}}
\end{center}
\end{table}
From Table~\ref{TJR}, it is clear that the method is readily capable
of identifying the number and position of poles even when the
associated resonances are broad and overlapping. 
The decrease in $\chi^2$ when going from no resonance to one resonance
is large, but far less significant than the fall when we ask is there
one or two resonances. 
From this Table, one would have no doubt that there are two resonances
in the model data and their positions are well determined. 

\begin{figure}[!htb]
\begin{center}
\includegraphics[width=0.9\textwidth,height=0.4\textheight]{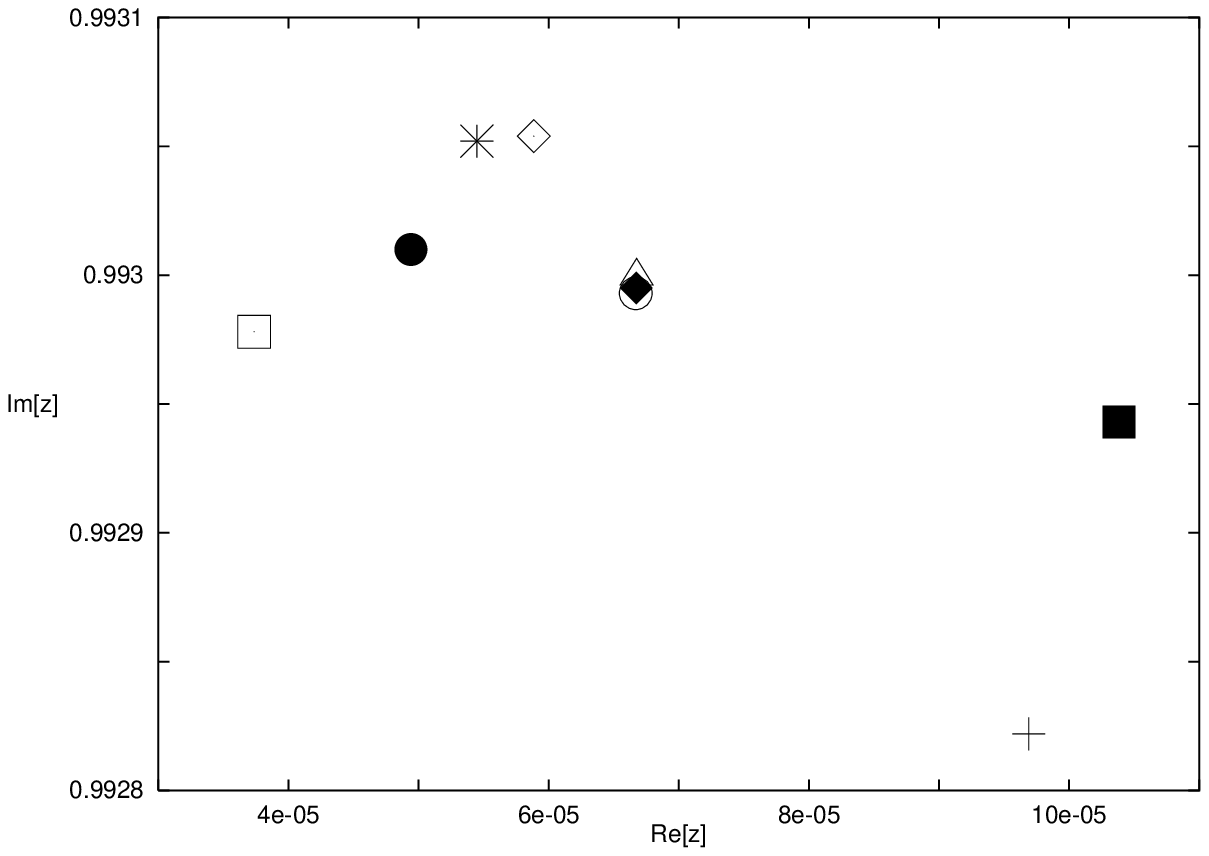}
\captive{Expanded view of the $z$-plane showing the spread of pole
positions found by varying the input parameters. 
$\blacklozenge$: standard values.
$\lozenge$: double the number of terms in $g(z)$.
$+$: halve the number of terms in $g(z)$.
$\ast$: Double the number of terms in $\chi^2$.
$\square$: Halve the number of terms in $\chi^2$.
$\circ$: Double the value of $f(z)$ in the unphysical region.
$\triangle$: Halve the value of $f(z)$ in the unphysical region.
$\blacksquare$:  Halve the errors in the unphysical region.
$\bullet$: Double the number of points in the unphysical region, i.e
not in the direct channel.
\label{zpos}}
\end{center}
\vspace{-5mm}
\end{figure}

Evaluating the errors on the pole positions that we obtain using this
method is nevertheless not straightforward. 
The experimental errors are folded into the actual calculation and so
their propagation cannot easily be followed. 
Moreover, as pointed out by Nogov\'a {\it et al.}~\cite{Nogova}, the
$\chi^2$'s obtained are not strictly statistical, so the standard
confidence level techniques are not appropriate.
Thus we resort to an order of magnitude estimate of the errors
obtained by varying the inputs that introduce uncertainty into the
calculation. 
These inputs include the number of terms in the Fourier expansion
defining the weight function, the number of terms in the summation
used to evaluate $\chi^2$ and the treatment of the unphysical region. 
The uncertainty in the pole position found in the $z$-plane due to
these changes in parameters is shown in Fig.~\ref{zpos}, from which we
see that the typical uncertainty in the $z$-plane is of the order of
$(4 + 20 i) \times 10^{-5}$.

\begin{table}[!htb]
\vspace{4mm}
\begin{center}
\begin{tabular}{|c|c|c|c|c|cc|}
\hline
Max.& $\chi^2_{0}$ & $\chi^2_{1}$ & $\chi^2_{2}$ & $\lambda $ & $\lambda_1$ & $\lambda_2 $\\
Energy  & & & & (MeV) &\multicolumn{2}{c|}{(MeV)}\\
\hline \hline
22.51 & 2320 & 43.3~ & 2.02 & $1395 \pm 53 i $ & $1938 \pm 234 i $ & $1453 \pm 137 i $\\
\hline
12.51 & 2361 & 37.0~ & 1.98 & $1394 \pm 56 i $ & $1951 \pm 253 i $ & $1456 \pm 133 i $\\
\hline
~8.51 & 2389 & 35.0~ & 1.96 & $1394 \pm 58 i $ & $1959 \pm 264 i $ & $1457 \pm 130 i $\\
\hline
~6.51 & 2380 & 18.9~ & 1.84 & $1395 \pm 69 i $ & $2024 \pm 331 i $ & $1458 \pm 114 i $\\
\hline
~3.51 & 2207 & 11.4~ & 1.77 & $1396 \pm 78 i $ & $2099 \pm 441 i $ & $1454 \pm 102 i $\\
\hline
~2.91 & 1959 & ~7.33 & 1.70 & $1398 \pm 84 i $ & $2178 \pm 788 i $ & $1445 \pm ~91 i $\\
\hline
~2.51 & 1529 & ~4.58 & 1.57 & $1399 \pm 92 i $ & $899 \pm 1457 i $ & $1422 \pm ~82 i $\\
\hline
\end{tabular}
\captin{$\chi^2$'s and pole positions for trial data described in the
text. 
Maximum energies are in~GeV. 
$\lambda$ is the pole position found when searching for one
resonance. 
$\lambda_1$ and $\lambda_2$ are the pole positions found when
searching for two resonances. 
$\chi^2_0$, $\chi^2_1$, and $\chi^2_2$ are the $\chi^2$'s assuming no,
one and two resonances respectively. 
The actual pole positions are ($1421 \pm 119 i$)~MeV for the
$\kappa_2$ and ($1957 \pm 106 i$)~MeV for the $\kappa_3$. 
\label{TBW}}
\end{center}
\vspace{-4mm}
\end{table}

Of course the physical amplitude will contain many resonances with
masses greater than those of our $\kappa_1$ and $\kappa_2$. 
As was stated earlier, this method is expected to be less sensitive to
these poles. 
To see to what extent this is true a further test was carried out
using trial data, up to various energies, containing an effective
range background, a $\kappa_2$ and another state at 1.95~GeV, which we
refer to as the $\kappa_3$. 
Data points were created at the same energies as the LASS data and
above that at 40~MeV intervals with an error of 5\%. 
The $\chi^2$'s and pole positions found are shown in Table~\ref{TBW}.
From these results we can clearly see that, as the maximum energy of
the data falls, the $\kappa_3$ becomes less necessary to describe the
data in the $z-$plane. 
With data up to 22.5~GeV, there is a strong case for claiming that
there are both the $\kappa_2$ and the $\kappa_3$, but their parameters
are not accurately found. 
When the data only extends to 2.5~GeV, the ratio of $\chi^2_1$ to
$\chi^2_2$ is small enough that one would not be confident in claiming
that the data exhibited more than one resonance. While the position of
the $\kappa_2$ is more stable, the parameters found for the $\kappa_3$
bear no resemblance to the correct values.
Meanwhile, the pole parameters found when testing for just one pole
are always recognisable as the $\kappa_2$.
If we introduce realistic experimental errors and inelasticity, then
the $\kappa_3$ becomes even more hidden. 

Due to the non-linearity of the mapping a fixed absolute uncertainty
in the $z$-plane will give different uncertainties in the $E$-plane
depending on the actual position. 
The minimum uncertainty in the position of the $\kappa_2$ is of the
order of 3~MeV on the real part and 5~MeV on the imaginary. 
For the $\kappa_3$ the minimum uncertainty would be 5~MeV on the real
part and 15~MeV on the imaginary part. 
In contrast, for the $\kappa_1$ the minimum uncertainty is of the
order 1~MeV on both parts. 
The pole positions listed in Table~\ref{TJR} would suggest that these
minimum uncertainties underestimate the true uncertainty and we can
expect to do no better than an accuracy of 10~MeV on the real part and
20~MeV on the imaginary part of the pole position. 
\vspace{-1mm}

\section{Results from Real Data}

\begin{figure}[!htb]
\begin{center}
\includegraphics[width=0.49\textwidth,height=0.35\textheight]{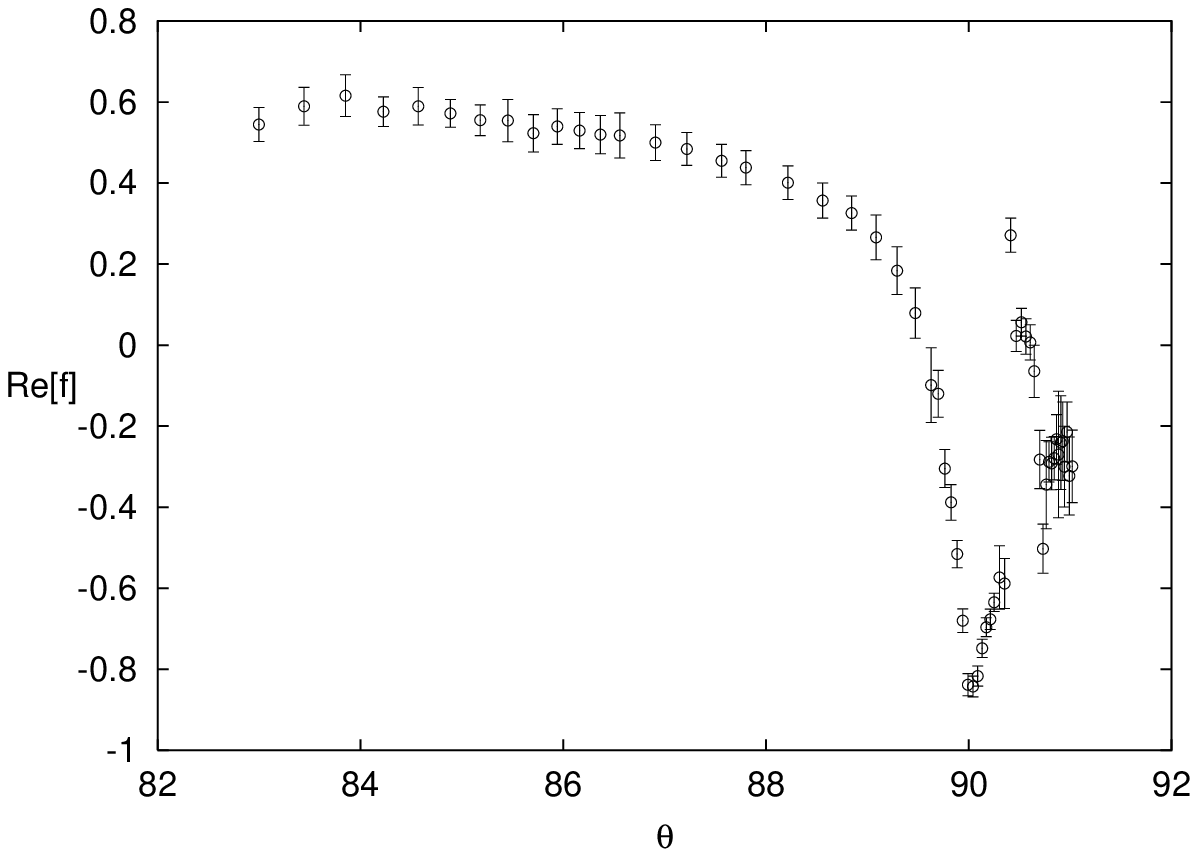}
\includegraphics[width=0.49\textwidth,height=0.35\textheight]{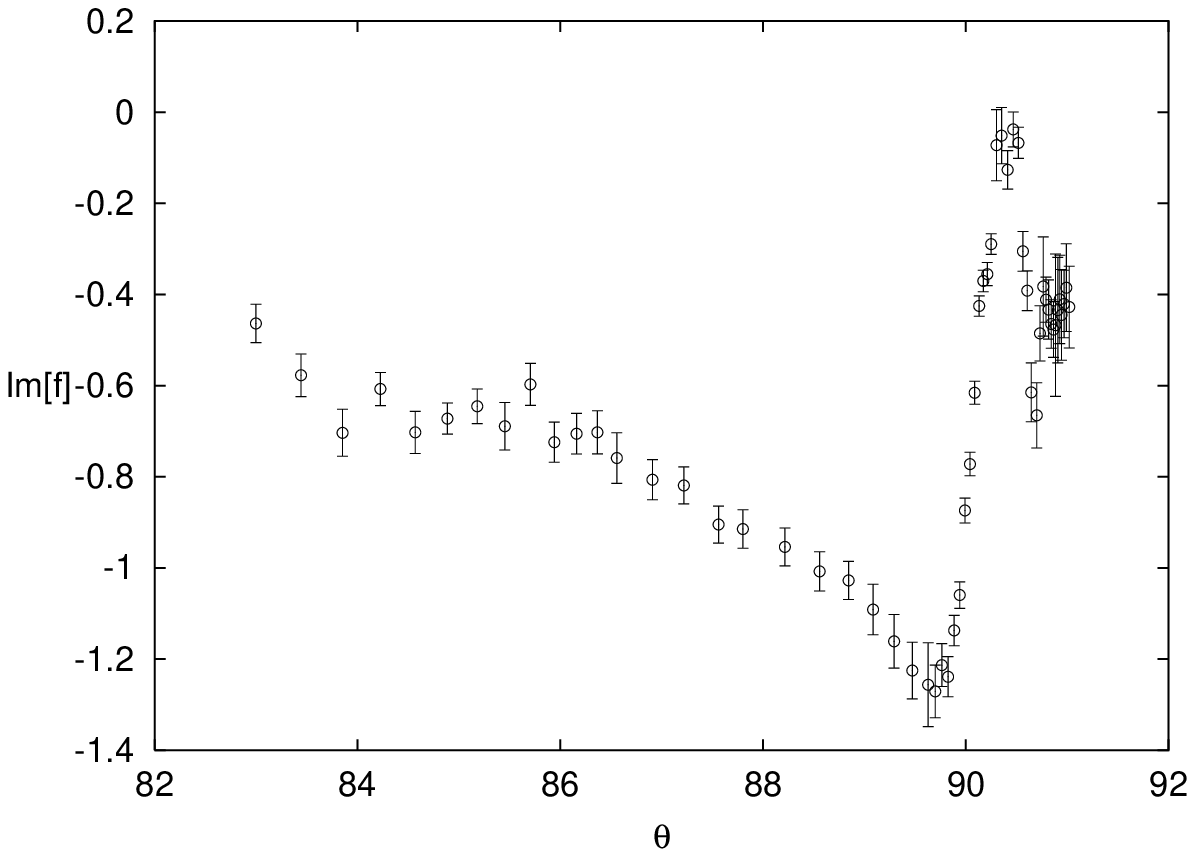}
\captive{Real and imaginary parts of $f^{II}$ as a function of
$\theta$ for the LASS $S-$wave data-set. 
Only the actual data points are shown, with unweighted errors.
\label{ladat}}
\end{center}
\end{figure}

\noindent Having assessed the capabilities and limitations of the method using
model data, we now apply it to the results from real experiments.
From LASS~\cite{Aston}, we have the magnitude and phase of the $\pi^+
K^-$ $S-$wave amplitude from 825~MeV up to 2.51~GeV as shown in
Fig.~\ref{ladat}. 
Due to a Barrelet ambiguity, the LASS group find two partial wave
amplitude solutions which only differ above 1.84~GeV. 
For our calculation we use their Solution A, but this choice of
solution does not affect our conclusions regarding the $\kappa(900)$
and the $K_0^*(1430)$.
The total $S-$wave amplitude for $\pi^+ K^-\to\pi^+ K^-$ is related to
the amplitudes with definite isospin\footnote{This being a physical process, the imaginary part of each partial wave amplitude must be positive. The authors of Ref.~\cite{Jamin} have noted (after the present work was complete) that this favours solution A of LASS~\cite{Aston}.} by 
\begin{equation}
f_S(s) = f^{(1/2)} + \frac{1}{2}\, f^{(3/2)}\qquad .
\end{equation}

Since resonances are only expected in the $I=1/2$ channel, it is
natural to consider the effect of separating out this component. 
Such a separation requires a modelling of the $I=3/2$ contribution.
Below 1.58~GeV, the LASS group provide a model of this contribution
based on the parametrisation of Estabrooks~\cite{Estabrooks2}.
This allows us to apply the method used in Sect.~4 to the $I=1/2$
amplitude alone, but only below 1.58~GeV. 
As a control on this procedure, we also consider the full $S-$wave
data over this reduced energy range, which we refer to as the ``Short
$S-$wave data-set''. 

The effective range type formula provided by LASS gives one possible
extrapolation of their data to threshold. 
This serves as {\it a guide} (and only a guide) to possible data
points between threshold and the start of the data. 
Note that the superscript $(I)$ is an isospin label.
All equations for amplitudes in this section refer to the physical
sheet. 
Their formula is
\begin{equation}
f^{(I)} = \frac{1}{\rho \, (\cot \delta^{(I)}_{BG} - i)} + \frac{e^{2
i \delta^{(I)}_{BG}}}{\rho \, (\cot \delta_{BW} - i)}\quad , 
\end{equation}

\noindent where the resonance term only appears in the $I={1}/{2}$ case and
\begin{eqnarray}
q \cot \delta^{(I)}_{BG} & = & \frac{1}{a^{(I)}} + \frac{1}{2}\,b^{(I)}\, q^2
\quad ,\label{er}\\
& & \nonumber \\
\cot \delta_{BW} & = & \frac{(m_r^2 - s)\,E \, q_r}{m_r^2\, \Gamma_r\,
q} \quad . 
\end{eqnarray}

\begin{sloppypar}
\noindent $q$ is given by Eq.~(\ref{mom}), and $q_r$ is its value at
$s=m_r^2$. 
The LASS fit\footnote{These parameters were provided by W. Dunwoodie
and are not the values quoted in the original paper.} gave
\mbox{$a^{({1}/{2})} = 2.19$~GeV$^{-1}$,} $b^{(1/2)} =
3.74$~GeV$^{-1}$, $a^{(3/2)} = -1.03$~GeV$^{-1}$, \mbox{ $b^{(3/2)} =
-0.94$~GeV$^{-1}$,} $m_r = 1.412\ $~GeV and $\Gamma_r = 0.294\ $~GeV. 
It is important to recognise that though the LASS effective range fit
assumes the tail of a Breit-Wigner-like pole, this does not prejudge
that such a pole exists in our analysis. 
We call this treatment of the low energy region \emph{Case A}.
\end{sloppypar}

While the above \emph{Case A} is largely experimentally motivated, a more
theoretically well-founded guide to low energy meson-meson scattering
is provided by Chiral Perturbation Theory ($\chi$PT). 
This makes predictions for near threshold $\pi K$ scattering that we
can input into our analysis. 
However, the effect of the higher order corrections becomes larger and
less immediately predictable as one goes much above threshold. 
Consequently, we only generate data points based on $\chi$PT within
100~MeV of threshold, with a precision encompassing the range of
present calculations~\cite{Bernard,Dobado,Jamin}. 
We call this \emph{Case B}.
We let the method determine the interpolation between the low energy
$\chi$PT results and where LASS data begin. 
This avoids the need for us to enter into any debate about whether
higher order corrections in $\chi$PT are summed better by the Inverse
Amplitude method~\cite{Dobado,Oller2} or by explicitly including
resonances~\cite{Bernard}. 
Such differences are large at 825~MeV.
It is in this sense that we describe our method as ``model
independent''.

\begin{figure}[!htb]
\begin{center}
\includegraphics[width=0.95\textwidth]{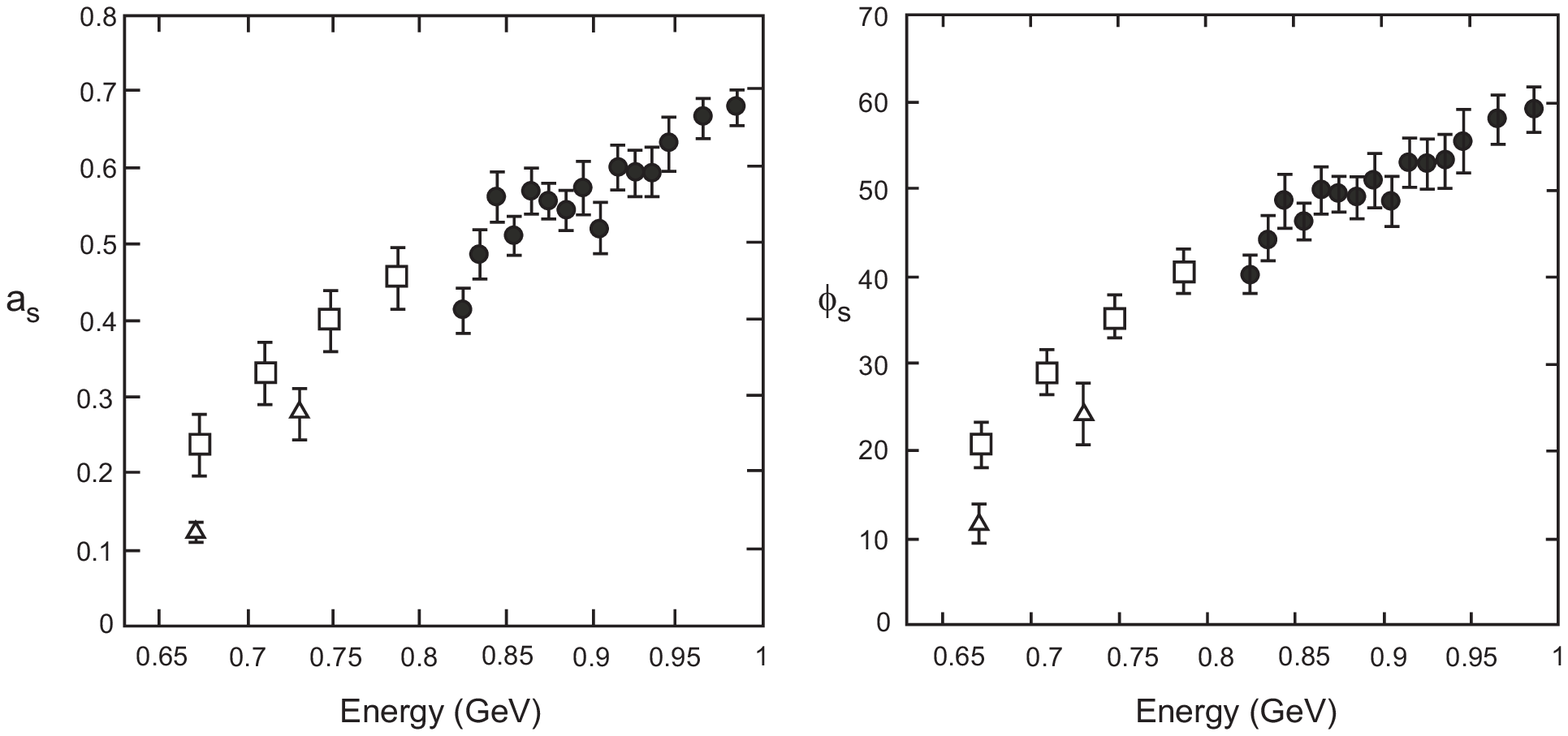}
\captive{Magnitude, $a_s$, and phase, $\phi_s$ (in degrees), of the
$\pi^+ K^-$ $S-$wave below 1~GeV from LASS data ($\bullet$),
contrasting the near threshold fit by the experimental group
($\square$) with the predictions of $\chi$PT ($\triangle$).
\label{lowdata}}
\end{center}
\end{figure}

In implementing either \emph{Case A} or \emph{Case B}, we only create
a few low energy ``data'' points so as not overly to prejudice the
results.
These two alternative sets of ``data'' are shown in Fig.~\ref{lowdata}
together with the LASS experimental results below 1~GeV.
In both Cases the unphysical cuts are treated as for the model
data-set of Sect.~4.

\vspace{-6mm}

\begin{table}[!htb]
\begin{center}
\begin{tabular}{|c|c|c|c|c|r|}
\hline
Case & Option & No. of  & $z_{pole}$ & $\sqrt{s_{pole}}$ & $\chi^2$~~ \\
& & resonances & $(r,\theta)$ & (MeV) &\\
\hline \hline
&  & 0 &  --- & --- & 1373~~ \\
\cline{3-6}
& 1 & 1 & $ (0.994,90.07^{\circ}) $ & $1433 \pm 149 i$ & 4.7\\
\cline{3-6}
&  & 2 & $ (0.995,90.15^{\circ}) $ & $1432 \pm 148 i$ & 1.8\\
A &  & & $ (0.987,82.02^{\circ}) $ & $~805 \pm ~13 i$ & \\
\cline{2-6}
&  & 0 & --- & --- & 1629~~ \\
\cline{3-6}
& 2 & 1 & $ (0.993,90.06^{\circ}) $ & $1423 \pm 157 i$ & 9.0\\
\cline{3-6}
&  & 2 & $ (0.993,90.06^{\circ}) $ & $1423 \pm 154 i$ & 8.8\\
&  & & $ (0.998,86.98^{\circ}) $ & $~969 \pm ~~6 i$ &\\
\hline
& & 0 & --- & --- & 1340~~ \\
\cline{3-6}
& 1 & 1 & $(0.994,90.10^{\circ})$ & $1444 \pm 151 i$ & 16.9\\
\cline{3-6}
& & 2 & $(0.994,90.07^{\circ})$ & $1436 \pm 133 i$ & 1.0\\
B & & & $(0.854,95.14^{\circ})$ & $~609 \pm 238 i$ & \\
\cline{2-6} 
& & 0 & --- & --- & 1694~~ \\
\cline{3-6}
& 2 & 1 & $(0.993,90.08^{\circ})$ & $1428 \pm 168 i$ & 33.2\\
\cline{3-6}
& & 2 & $(0.994,90.05^{\circ})$ & $1429 \pm 132 i$ & 1.0\\
& & & $(0.870,93.35^{\circ})$ & $~645 \pm 260 i$ &\\
\hline
\end{tabular}
\captin{Pole positions and $\chi^2$'s for LASS $S-$wave
data~\cite{Aston} from 0.825~GeV to 2.51~GeV.  
Cases A and B are explained in the text and shown in Fig.~\ref{lowdata}.
\newline Option 1: The amplitude in the unphysical region equals the
amplitude at threshold, with an error of 5. 
\newline Option 2: The amplitude in the unphysical region equals the
amplitude at threshold, with an error of 2.5.
\label{TLSLR}}
\end{center}
\vspace{-5mm}
\end{table}

In Table~\ref{TLSLR}, we present the results of our analysis for the
data shown in Fig.~\ref{ladat} and the two extrapolations shown in
Fig.~\ref{lowdata}. 
The Table shows the pole positions found for the full $S-$wave data,
with full and halved errors on the unphysical points and for the
\emph{Case A} and \emph{Case B} continuations to threshold.
It is worth pointing out that our method will `find' exactly as many
poles as it is asked to.
So if we search for two resonances then positions for two resonances
will be given no matter how many resonances are present in the data. 
Resonances that really are present in the data will be stable to
changes in unphysical parameters and result in sizeable falls in the
$\chi^2$. 
Conversely, if the $\chi^2$ does not fall significantly between one
and two resonances then we can conclude that the second resonance
`found' is not really present and the pole position given is
meaningless. 
Likewise, any resonance found whose pole position changes wildly with
variations in parameters not determined explicitly by the experimental
$\pi K$ data used will also be an artifact.  

\baselineskip=6.2mm

In order to corroborate the results shown in Table~\ref{TLSLR} the
results for the $I = 1/2$ and the shorter $S-$wave data-sets,
using the \emph{Case A} continuation to threshold are shown in
Tables~\ref{TLIR} and~\ref{TLSSR}.
These Tables show a remarkable consistency, as one would hope for
effects that are real. 

\begin{table}[!htb]
\begin{center}
\begin{tabular}{|c|c|c|c|r|}
\hline
Option & No. of & $z_{pole}$ & $\sqrt{s_{pole}}$ & $\chi^2$~~ \\
& resonances & $(r,\theta)$ & (MeV) & \\
\hline \hline
& 0 & --- & --- & 210~~ \\
\cline{2-5}
1 & 1 & (0.992,90.22$^o$) & $1467 \pm 224 i$ & 2.1\\
\cline{2-5}
& 2 & (0.992,90.22$^o$) & $1464 \pm 221 i$ & 1.1\\
& & (0.993,82.14$^o$) &$~808 \pm ~~7 i$ & \\
\hline
& 0 & --- & --- & 367~~ \\
\cline{2-5}
2 & 1 & (0.991,90.23$^o$) & $1455 \pm 245 i$ & 5.3\\
\cline{2-5}
& 2 & (0.991,90.19$^o$) & $1439 \pm 237 i$ & 2.4\\
& & (0.911,100.1$^o$) & $503 \pm 238 i$ & \\
\hline
\end{tabular}
\captin{Pole positions and $\chi^2$'s for LASS $I=\frac{1}{2}$
data~\cite{Aston}. 
\newline Option 1 has unphysical errors set to 5.
Option 2 has unphysical errors set to 2.5.
\label{TLIR}}
\end{center}
\vspace{-3mm}
\end{table}

\begin{table}[!htb]
\begin{center}
\vspace{5mm}
\begin{tabular}{|c|c|c|c|r|}
\hline
Option & No. of  & $z_{pole}$ & $\sqrt{s_{pole}}$ & $\chi^2$~~ \\
& resonances & $(r,\theta)$ & (MeV) & \\
\hline \hline
& 0 & --- & --- & 170~~ \\
\cline{2-5}
1 & 1 & (0.991,90.03$^o$) & $1471 \pm 242 i$ & 2.2\\
\cline{2-5}
& 2 & (0.985,90.07$^o$) & $1422 \pm 553 i$ & 1.9\\
& & (0.997,90.08$^o$) & $1457 \pm ~80 i$ & \\
\hline
& 0 & --- & --- & 283~~ \\
\cline{2-5}
2 & 1 & (0.991,90.26$^o$) & $1459 \pm 264 i$ & 5.6\\
\cline{2-5}
& 2 & (0.990,90.23$^o$) & $1440 \pm 263 i$ & 2.4\\
& & (0.886,101.8$^o$) & $~513 \pm 208 i$ &\\
\hline
\end{tabular}
\captin{Pole positions and $\chi^2$'s for the short LASS $S-$wave
data~\cite{Aston}. 
\newline Option 1 has unphysical errors set to 5.
Option 2 has unphysical errors set to 2.5.
\label{TLSSR}}
\end{center}
\vspace{-7mm}
\end{table}

\begin{figure}[p]
\begin{center}
\includegraphics[width=0.95\textwidth,height=0.35\textheight]{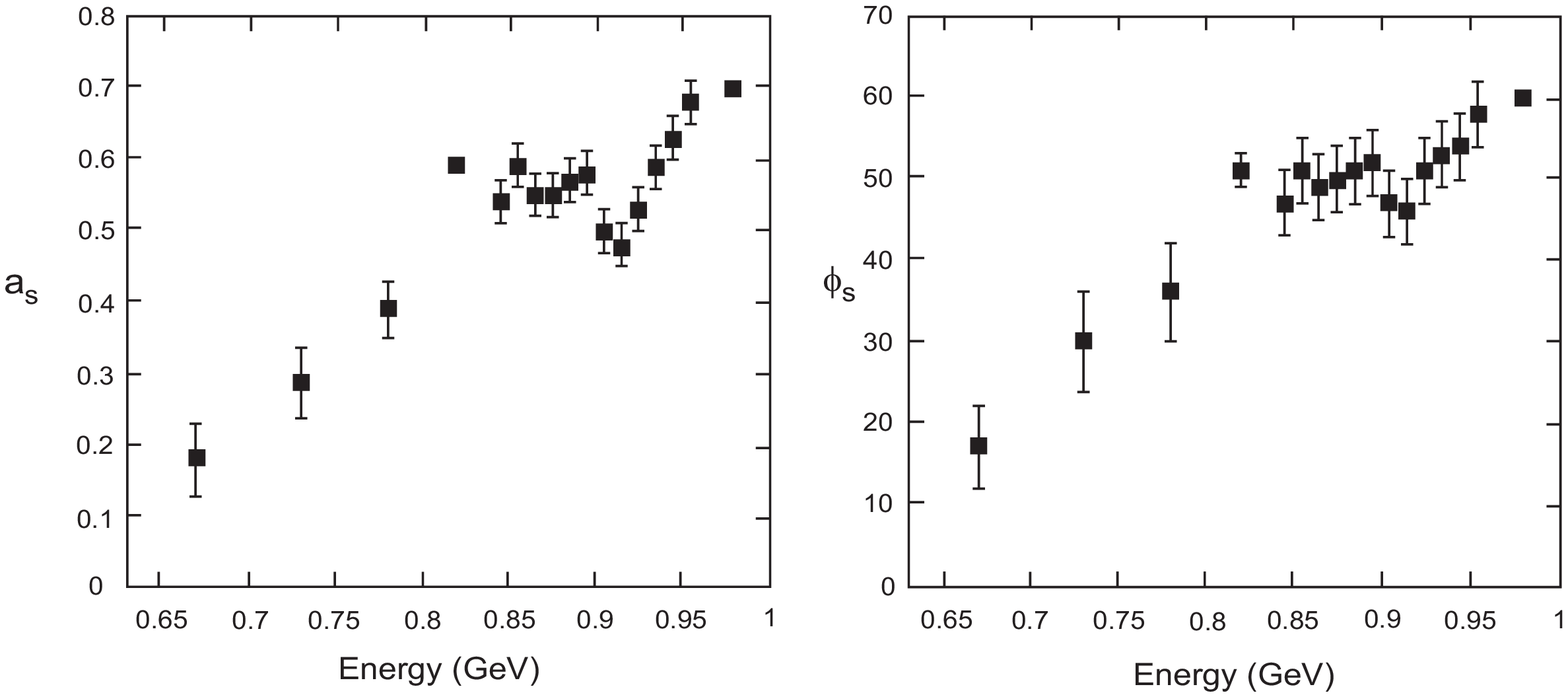}
\captive{Magnitudes, $a_s$, and phases, $\phi_s$ (in degrees)  $S-$wave $\pi K$ scattering amplitude below 1~GeV from Estabrooks
\emph{et al.}~\cite{Estabrooks}, to be compared with Fig.~6.
\label{esta}}
\end{center}
\end{figure}

\begin{table}[p]
\begin{center}
\begin{tabular}{|c|c|c|c|c|r|}
\hline
Case & Option & No. of & $z_{pole}$ & $\sqrt{s_{pole}}$ & $\chi^2$~\\
& & resonances & $(r,\theta)$ & (MeV) & \\
\hline \hline
& & 0 & --- & --- & 553\\
\cline{3-6}
& 1 & 1 & (0.991,90.20$^o$) &$1446 \pm 232 i$ & 16\\
\cline{3-6}
& & 2 & (0.986,89.83$^o$) &$1277 \pm 231 i$ & 12\\
A & & & (1.017,90.46$^o$) &$1344 \pm 444 i$  & \\
\cline{2-6}
& & 0 & --- & --- & 812\\
\cline{3-6}
& 2 & 1 & (0.990,90.16$^o$) & $1411 \pm 250 i$ & 14\\
\cline{3-6}
& & 2 & (0.988,89.92$^o$) & $1323 \pm 221 i$ & 14\\
& & & (1.01,90.47$^o$) & $ 1564 \pm 316 i$ & \\
\hline
& & 0 & --- & --- & 551\\
\cline{3-6}
& 1 & 1 & $(0.991,90.20^{\circ})$ & $1447 \pm 231 i$ & 17\\
\cline{3-6}
& & 2 & $(0.986,89.84^{\circ})$ & $1286 \pm 227 i$ & 14\\
B & & & $(1.015,90.44^{\circ})$ & $1382 \pm 423 i$ & \\
\cline{2-6}
& & 0 & --- & --- & 816\\
\cline{3-6}
& 2 & 1 & $(0.990,90.16^{\circ})$ & $1411 \pm 253 i$ & 16\\
\cline{3-6}
& & 2 & $(0.991,89.81^{\circ})$ & $1324 \pm 159 i$ & 16\\
& & & $(1.001,89.89^{\circ})$ & $1395 \pm ~18 i$ & \\
\hline 
\end{tabular}
\captin{Pole positions and $\chi^2$'s for Estabrooks {\it et
al.}~\cite{Estabrooks} $S-$wave data. The Cases A and B are described
in the text.
\newline  Option 1 has unphysical errors set to 5.
Option 2 has unphysical errors set to 2.5.
\label{TER}}
\end{center}
\vspace{-1mm}
\end{table}

As a further check of our results, the same technique was applied to
the total $S-$wave data from an earlier experiment~\cite{Estabrooks}. 
These data extend closer to threshold, so it is only necessary to
create a point at threshold.
Barrelet ambiguities lead to four possible partial-wave solutions,
which differ only above 1500~MeV, so the choice of solution should not
affect our conclusions regarding the $\kappa$.
We have carried out our analysis using solutions A and B and found the
results to be qualitatively similar and so we show only those for
solution~B.
The data are shown below 1~GeV in Fig.~\ref{esta}, which should be
compared with the low energy treatment of the LASS data-set (see
Fig.~\ref{lowdata}).
This new \emph{Case A} is defined using an effective range formula,
see Eq. \ref{er}, with $a^{(1/2)} = 2.39$~GeV$^{-1}$ and $a^{(3/2)} =
-1.00$~GeV$^{-1}$.
No resonance was assumed in this fit, so any lingering doubts that
assuming the tail of the $K_0^*(1430)$ in the low energy extrapolation
introduces a prejudice is clearly not there in this case.
We also consider \emph{Case B} using the scattering length predictions
of $\chi$PT, as before.
The results for the Estabrooks {\it et al.} data-set~\cite{Estabrooks}
are shown in Table~\ref{TER}.

For the real $\pi K$ experimental results,
Tables~\ref{TLSLR}--\ref{TER} display a consistency in identifying a
single resonance which looks very like the $K_0^*(1430)$. 
The fall in $\chi^2$ in going from no resonance to one resonance is
always sizeable, ranging from a factor of 35 to a factor of 290. 
In contrast when going from one to two resonances, the $\chi^2$ does
not always fall by a significant amount.
Recall importantly that the present analysis technique is always most
sensitive to the lightest resonance and this is clearly the
$K_0^*(1430)$.
So whilst the $K^*_0(1430)$ is overwhelmingly apparent, a lighter,
broad resonance in the LASS range above 825~MeV is not.
When using the LASS data and \emph{Case B}, where the low energy
behaviour is given by $\chi$PT, there is a sizeable decrease in
$\chi^2$ for two resonances, but the very light and broad one is below
the experimental ranges studied by both LASS and Estabrooks \emph{at
al.}.

Halving the errors in the unphysical region should not greatly affect
the pole positions obtained, and from Table~\ref{TJR} we can see that
this is indeed the case. 
From Tables~\ref{TLSLR}--\ref{TER} we see that for the assumption that
there is only one resonance present, then halving the unphysical
errors or changing the near threshold continuation, has a small effect. 
However, when we force the amplitude to have two resonances the effect
is much more noticeable. 
Generally, the resonance which looks like $K_0^*(1430)$ tends to stay
in a similar position, but the second one moves around wildly. 
This is what one would expect if this second resonance is merely an
artifact of forcing the method to find two poles that are not really
present in the energy range of the data that are being analytically
continued into the complex plane.

We mentioned earlier that we would expect the method to be less
sensitive to high mass resonances. 
This is borne out by these results.
The LASS group provide strong evidence for a state at 1.95~GeV.
With data to 2.51~GeV we would expect to see this state showing up,
but the compression of the high energy portion of the amplitude has
diluted the effect of the $K_0^*(1950)$ so much that it is not needed
to describe the data in the complex $z-$plane. 
This is to be expected from our study using model data, described in
Sect.~4. 
The mapping procedure severely limits our ability to find states above
1800~MeV in $\pi K$ scattering. 
However, below that, and particularly down towards the lower end of
the experimental range, the method is totally reliable and quite
unambiguous. 
Consequently, we have no doubt that
\vspace{-1mm}
\begin{itemize}
\item[i.]  there is a $K^*_0(1430)$,
\item[ii.] there is no $\kappa(900)$,
\end{itemize}
in experimental $\pi K$ scattering data.

\section{Conclusions}
We have here applied a method of analytic continuation in as model independent way as is presently possible to the existing
data on $\pi^+K^-$ scattering between 825~MeV and 2.5~GeV. We find
that there is only one scalar resonance between 825 and 1800~MeV,
which is readily identified with the $K^*_0(1430)$. 
Our procedure requires no assumptions about the amplitude being
described by Breit-Wigner forms with any particular form of
background. 
It directly counts the number of poles of the $S-$matrix on the nearby
unphysical sheet. 
This method provides a rigorous test of whether the $\kappa(900)$
exists, and we conclude it does not. 
This is so whether we use the older Estabrooks {\it et
al.}~\cite{Estabrooks} data or that of the LASS
collaboration~\cite{Aston}.
Clearly, the question of the existence of a $\kappa$ of mass and width
that does not intrude into the energy range explored by experiment
cannot be addressed.

The fact that there is only one strange scalar between 800 and
1800~MeV has implications for quark models that postulate both a
$q{\overline q}$ and a $qq{\overline {qq}}$ nonet in this
region~\cite{Black2,Alford}.
Our results of course do not depend on this modelling.
However, we cannot refrain from commenting that we believe that other
calculations clearly show the scalars are different from other states
with underlying $q{\overline q}$ composition. 
Their decay modes, overwhelmingly to two mesons, mean that the
physical hadrons do have a large $qq{\overline {qq}}$ component in
their Fock space~\cite{VanBeveren,Oller1,Tornqvist,Boglione}. 
Thus for many purposes the scalars below 1800~MeV (particularly those
with flavour) do behave like 4--quark states, even though they may
well be seeded by single $q{\overline q}$ pairs at the bare level. 
However, such calculations and speculations are for elsewhere.

We conclude that the data on $S-$wave $\pi K$ scattering exhibit just
one pair of complex conjugate poles between 0.8 and 1.8~GeV. 
This means that there is only one resonance present in this channel in
the region accessed by the presently highest statistics experiment.
This resonance has a mass around 1400~MeV and a width of about 300~MeV.
The $\kappa(900)$ does not exist.

\section*{Acknowledgements}
We are most grateful to Bill Dunwoodie for supplying us with the LASS
data and a suitable parametrisation down to threshold. 
This work was supported in part under the  EU-TMR Programme, Contract
No. CT98-0169, EuroDA$\Phi$NE. 
One of us (S.N.C.) acknowledges receipt of a studentship from the
U.K. EPSRC.


\begin{thebibliography}{99}

\bibitem{Groom}
D.~E.~Groom {\it et al.},
Eur.\ Phys.\ J.\  {\bf C15} (2000) 1.

\bibitem{Jaffe1}
R.~Jaffe,
Phys.\ Rev.\  {\bf D15} (1977) 267;\\
Phys.\ Rev.\  {\bf D15} (1977) 281.

\bibitem{Weinstein}
J.~Weinstein and N.~Isgur,
Phys.\ Rev.\ Lett.\  {\bf 48} (1982) 659.

\bibitem{glueballs} see, for instance, E. Klempt,
Acta. Phys. Polon. {\bf B29} (1998) 3367;\\ 
F.E. Close and P.R. Page, Sci. Am. {\bf 279} (1998) 52;\\
C. Michael, hep-ph/9810415, Nucl. Phys. {\bf A655} (1999) 12;\\
M.R. Pennington, hep-ph/9811276, Proc. of {\it Workshop on Photon
Interactions and the Photon Structure}, Lund, 1998 (ed. G. Jarlskog
and T. Sj\"ostrand) pp.~313-328.

\bibitem{Black2}
D.~Black, A.~H.~Fariborz, F.~Sannino and J.~Schechter,
Phys.\ Rev.\  {\bf D59} (1999) 074026;\\
D.~Black, A.~H.~Fariborz and J.~Schechter,
Phys.\ Rev.\  {\bf D61} (2000) 074001.

\bibitem{Jaffe2}
R.~L.~Jaffe,
hep-ph/0001123.

\bibitem{Scadron}
M.~D.~Scadron,
Phys.\ Rev.\  {\bf D26} (1982) 239.

\bibitem{Delbourgo}
R.~Delbourgo and M.~D.~Scadron,
Int.\ J.\ Mod.\ Phys.\  {\bf A13} (1998) 657.

\bibitem{Black1}
D.~Black, A.~H.~Fariborz, F.~Sannino and J.~Schechter,
Phys.\ Rev.\  {\bf D58} (1998) 054012.

\bibitem{Ishida}
S.~Ishida, M.~Ishida, T.~Ishida, K.~Takamatsu and T.~Tsuru,
Prog.\ Theor.\ Phys.\  {\bf 98} (1997) 621.

\bibitem{VanBeveren}
E.~Van Beveren, T.~A.~Rijken, K.~Metzger, C.~Dullemond, G.~Rupp and J.~E.~Ribeiro,
Z.\ Phys.\  {\bf C30} (1986) 615.

\bibitem{Oller1}
J.~A.~Oller and E.~Oset,
Phys.\ Rev.\  {\bf D60} (1999) 074023.

\bibitem{Tornqvist}
N.~A.~Tornqvist,
Z.\ Phys.\  {\bf C68} (1995) 647.

\bibitem{Anisovich}
A.~V.~Anisovich and A.~V.~Sarantsev,
Phys.\ Lett.\  {\bf B413} (1997) 137.

\bibitem{Aston}
D.~Aston {\it et al.} (LASS Collaboration),
Nucl.\ Phys.\  {\bf B296} (1988) 493.

\bibitem{Estabrooks}
P.~Estabrooks, R.~K.~Carnegie, A.~D.~Martin, W.~M.~Dunwoodie, T.~A.~Lasinski and D.~W.~Leith,
Nucl.\ Phys.\  {\bf B133} (1978) 490.

\bibitem{Nogova}A.~Nogov\'a, J.~Pi\v s\'ut, and P.~Pre\v snajder, Nucl.\ Phys.\ {\bf B61}, (1973) 438;\\
A.~Nogov\'a, J.~Pi\v s\'ut, Nucl.\ Phys.\ {\bf B61}, (1973) 445.

\bibitem{Taylor}See e.g. J.~T.~Taylor {\it Scattering Theory} John Wiley \& Sons (1972);\\
R.~J.~Eden, P.~V.~Landshoff, D.~I.~Olive and J.~C.~Polkinghorne {\it The Analytic S-Matrix} CUP (1966).

\bibitem{Estabrooks2}
P.~Estabrooks,
Phys.\ Rev.\  {\bf D19} (1979) 2678.

\bibitem{Jamin}
M.~Jamin, J.~A.~Oller and A.~Pich,
hep-ph/0006045.

\bibitem{Bernard}
V.~Bernard, N.~Kaiser and U.~G.~Meissner,
Nucl.\ Phys.\  {\bf B357} (1991) 129;\\
V.~Bernard, N.~Kaiser and U.~G.~Meissner,
Nucl.\ Phys.\  {\bf B364} (1991) 283.

\bibitem{Dobado}
A.~Dobado and J.~R.~Pelaez,
Phys.\ Rev.\  {\bf D47} (1993) 4883.


\bibitem{Oller2}
J.~A.~Oller, E.~Oset and J.~R.~Pel\'aez,
Phys.\ Rev.\ Lett.\  {\bf 80} (1998) 3452;\\
J.~A.~Oller, E.~Oset and J.~R.~Pel\'aez,
Phys.\ Rev.\  {\bf D59} (1999) 074001;\\
\emph{Erratum-ibid.}\ {\bf D60} (1999) 099906

\bibitem{Alford}
M.~Alford and R.~L.~Jaffe,
hep-lat/0001023.

\bibitem{Boglione}
M.~Boglione and M.~R.~Pennington, hep-ph/9703257, Phys. Rev. Lett. {\bf 79} (1997) 1998.
\end{thebibliography}
\end{document}